[1994/12/01]

\documentclass[]{amsart}
\usepackage{amscd}
\usepackage{amsthm} 
\usepackage{amsfonts}
\usepackage{amssymb}
\usepackage{amsmath}
\usepackage{verbatim}

\newcommand{\labelrightarrow}[3]{\begin{CD}#1@>#2>>#3\end{CD}}

\begin{document}

\title[QCI]{Quantum Circuit Identities}

\author{ Chris Lomont}


\thanks{Research supported by AFRL grant F30602-03-C-0064}

\email{clomont@cybernet.com, clomont@math.purdue.edu}

\urladdr{www.math.purdue.edu/$\sim$clomont}

\curraddr{Department of Mathematics\\
Purdue University\\
150 N. University Street\\
West Lafayette, IN, 47907-2067}

\date{July 2, 2003}



\begin{abstract}
There are quantum circuit identities that simplify quantum
circuits, reducing the effort needed physically to implement them.
This paper constructs all identities made from 3 or fewer
operations taken from a common set of one qubit operations, and
explains how they may be used to simplify the cost of constructing
quantum circuit identities.
\end{abstract}

\maketitle

\section{Introduction}
Since the breakthrough algorithms of Grover \cite{Grover96} and
Shor \cite{Shor94}, their generalizations to the hidden subgroup
problem (HSP) by Kitaev \cite{Kitaev95}, and beyond
\cite{Hallgren00,Hallgren03}, there has been intense interest in
building quantum computers. But building quantum circuits is a
challenging task, and one important part is being able to reduce a
circuit to simple building blocks that are easily implemented.
Some blocks are more costly to implement in terms of time, effort,
and cost, for different physical realizations of quantum gates, so
exchanging some operations for equivalent operations can be
useful. Some different physical approaches are Raman-coupled
low-energy states of trapped ions \cite{Cirac95,Nagerl95},
electron spins in quantum dots \cite{Imam1999}, linear optics
\cite{Knill01}, nuclear spins in silicon \cite{Kane98}, and atomic
cavity quantum electrodynamics \cite{Rausch99}. Furthermore, when
building quantum simulators \cite{Hayes03,Bettelli01}, it is
useful to simplify constructions to equivalent circuits using
fewer or faster operations. A slightly old list of quantum
simulators is at \cite{SimulatorList}.

Often placing two larger circuit blocks in series allows the
adjacent blocks to have some gates merged. This can be done by
realizing when certain combinations of elementary operations can
be replaced by equivalent operations which lower the cost metric.
For a good introduction to quantum circuits see \cite{BBCD95}.

Identities can be generated automatically with reasonable effort,
but when automatically reducing circuits, doing so at runtime is
too slow for longer identities. It is therefore important to have
on hand a large list of identities, so simplifications can be
automated. The purpose of this paper is to start with a relatively
flexible set of matrices, and construct all identities up to a
given length from them. The value of this paper is a simple set of
rules to simplify many quantum circuits automatically and
deterministically. This is done explicitly for one qubit
identities made from products of 3 or less ``elementary'' gates.
Since most simplifications in practice should come from merging a
few gates in adjacent larger circuit blocks, choosing length 3
should encompass most needs.

This paper is laid out as follows: section \ref{s:definitions}
defines the matrices used in the identities and their relevance,
section \ref{s:procedure} explains the procedure to obtain the
identities (filtering out trivial or redundant identities, and
proving their validity), section \ref{s:identities} contains a
list of identities, and finally section \ref{s:conclusion}
contains concluding remarks.

\section{Definitions}\label{s:definitions}
Single qubit operations are in one-to-one correspondence with
$2\times 2$ unitary matrices, but certain of these matrices have
special importance in the quantum computing literature
\cite[Chapter 4]{ChuangNielsen}. Most circuits in the literature
are constructed from the following eleven unitary matrices:

$$\begin{array}{llll}
I=\left(\begin{array}{rr}1&0\\0&1\end{array}\right) &
X=\left(\begin{array}{rr}0&1\\1&0\end{array}\right) &
Y=\left(\begin{array}{rr}0&-i\\i&0\end{array}\right) &
Z=\left(\begin{array}{rr}1&0\\0&-1\end{array}\right) \\
S=\left(\begin{array}{rr}1&0\\0&i\end{array}\right) &
T=\left(\begin{array}{rr}1&0\\0&e^\frac{i\pi}{4}\end{array}\right)&
H=\frac{1}{\sqrt{2}}\left(\begin{array}{rr}1&1\\1&-1\end{array}\right)&
P(\theta)=e^{i\theta/2}I\\
\end{array}$$


$$\begin{array}{lrr}
R_x(\theta)=\cos\frac{\theta}{2}I-i\sin\frac{\theta}{2}X&
\;\;\;\;\;\;\;\;\;\;&
R_y(\theta)=\cos\frac{\theta}{2}I-i\sin\frac{\theta}{2}Y
\end{array}$$
$$R_z(\theta)=\cos\frac{\theta}{2}I-i\sin\frac{\theta}{2}Z$$


$X$, $Y$, and $Z$ are the Pauli spin matrices, and $R_x$, $R_y$,
and $R_z$ are the corresponding exponential matrices, giving
rotations on the Bloch sphere \cite[Exercise 4.6]{ChuangNielsen}.
$H$ is the Hadamard matrix, used extensively in quantum
algorithms. $S$ is the phase gate, and $T$ is called the $\pi/8$
gate (even though $\pi/4$ appears in it). Trivially $S=T^2$ and
$Z=S^2$; they are each included since they appear in many quantum
circuit constructions. $P$ is a phase rotation by $\theta/2$ to
help match identities automatically.

\section{Procedure}\label{s:procedure}
Infinite families of qubit operations are likely to be hard to
construct physically, except in special cases, so we arbitrarily
chose a \emph{finite} subset of the matrices in section
\ref{s:definitions} - small enough so that automatically
generating identities is not too time consuming, yet large enough
to yield a useful set of identities. We start with 35 matrices
$\Lambda=\{I,X,Y,Z,H,S,T,R_x(\theta_j),R_y(\theta_j),R_z(\theta_j),P(\theta_j)\}$
where $\theta_j=\frac{4\pi}{8}j$, for $j=1,2,\dots,7$. Note that
$R_x$, $R_y$, $R_z$, and $P$ all have period $4\pi$ and equal the
identity when their argument is 0, thus the choices for
$\theta_j$. Taking fewer choices for $\theta_j$ results in much
fewer identities, but enlarging the number by a factor of 2 made
the process unbearably slow. In order to shorten notation later,
let $X_j=R_x(\frac{4\pi}{8}j)$, $j=1,2,\dots,7$. Similarly define
$Y_j$, $Z_j$, and $P_j$.

The process to find the identities was programmed in both C++ and
Mathematica 4.2 to help weed out errors and avoid missing any
identities. An identity is an equivalence $lhs=rhs$, for example
$Z=HXH$, where $lhs$ and $rhs$ are strings of matrix names. The
length of an identity is the number of matrices on the right hand
side. $lhs$ always has length 1. To find all identities of length
$n$, all $n^{|\Lambda|}$ products of combinations of matrices from
$\Lambda$ are computed, and then the results are compared
numerically with a small tolerance to each matrix in $\Lambda$.
Numerical computations were a lot faster than doing accurate
symbolic comparisons in Mathematica, and they were much easier to
implement on the C++ side. This generated a list of identities of
the form $lhs=rhs$ where $lhs$ is a single matrix, and $rhs$ is a
product of n matrices. Identities were passed through a filter
(described in section \ref{s:filter}) to remove redundant
identities, trivial identities, and to clean them up by applying
obvious simplifications. Finally, the identities of length up to 3
are listed in section \ref{s:identities}, modulo the filtering
process. The final list of identities were verified symbolically
in Mathematica 4.2 \cite{Mathematica} and numerically by the C++
program to ensure that filtering did not introduce errors. C++
source code and a Mathematica 4.2 notebook implementing the
programs are at \cite{LomontCodeQC}, along with a file containing
all (unfiltered) identities of length 4 and less, of which there
are 400089 (see table \ref{t:counts}).

\subsection{Filtering identities}\label{s:filter}
Each identity is run through the filtering process described below
to remove redundant identities and those that can be shortened
using previous identities. The filter performs the following steps
on an identity:

{\begin{center}{\textbf{The Filter}}\end{center}}
\begin{enumerate}
\item \emph{Shrink}: If the length $d$ of $rhs$ is 3 or larger,
remove the identity if any identity with a length of
$2,3,\dots,d-1$ can be applied to shorten $rhs$. This way, knowing
$I=HH$, we avoid identities like $H=HHH$ which are deduced from
shorter identities.

\item \label{i:negate} \emph{Negate}: Apply the rules
$\{Q_4\rightarrow -I, Q_j\rightarrow -Q_{j-4}\}$ for
$Q\in\{P,X,Y,Z\}$ and $j=5,6,7$. This replaces expressions like
$X_5Y_6\rightarrow (-X_1)(-Y_2)=X_1Y_2$, making all subscripts
less than 4, and shortening the list.

\item \emph{Clean}: Merge the negative signs from the previous
step, and move to the front of the string. This step is trivial
mathematically, but must be performed symbolically in the
computer.

\item \emph{Phase}: Since the $P_j$ commute with all matrices,
merge them all to the front of $rhs$, except for a possible ``$-$"
in front. That is, replace expressions like $P_2=-P_3SP_1Z_1$ with
$I=-P_2SZ_1$.

\item \emph{Normalize}: Apply the 90 commuting rules
$\{QQ_j\rightarrow Q_jQ, TZ\rightarrow ZT, SZ\rightarrow ZS,
ST\rightarrow TS, SZ_j\rightarrow Z_jS, TZ_j\rightarrow Z_jT,
VQ_4\rightarrow Q_4V\}$ for $Q\in\{X,Y,Z\}$, $V\in\Lambda$, and
the rules on identities $\{II\rightarrow I,I\rightarrow\}$ to
normalize the expression, and to allow further simplifications.
For example, this allows the simplifications
$ZZ_3SZSZ_2\rightarrow ZZSZ_3Z_2\rightarrow SZ_5\rightarrow
-SZ_1$.

\item \label{i:collapse} \emph{Collapse}: Combine any obvious
expressions using the 196 identities $\{Q_iQ_j\rightarrow Q_k\}$
for $Q\in\{P,X,Y,Z\}$, $i,j\in\{1,2,\dots 7\}$, and $k=(i+j)\mod
8$, where $Q_0=I$. For example, $X_2X_3$ becomes $X_5$ and
$Y_2Y_3Y_3$ becomes $I$.

\item \emph{Merge}: Remove trivial identities such as when the
$lhs$ and $rhs$ are symbolically identical, or $lhs=I$. Remove
duplicate identities at this step.

\item \label{i:rotation} \emph{Rotate}: To avoid a lot of
identities, if the identity is now made up only of rotations
$X_j$, $Y_j$, and $Z_j$, (and treating $\pm I$ as rotations) it is
discarded. Since these matrices act as rotations of the Bloch
sphere they satisfy usual rotation identities, so this does not
remove too many useful relations. For the online list
\cite{LomontCodeQC} they are retained for completeness.

\item \emph{Repeat}: Repeat the above steps until the list of
length $n$ identities becomes stable. Repetition is necessary; for
example step \ref{i:collapse} may allow step \ref{i:negate} to
simplify the identity further.

\item \label{i:grouping} \emph{Grouping}: After the list is
stable, group similar identities to shorten the list, using
symbols $\{A,B,C\}$, chosen cyclically from $\{X,Y,Z\}$. For
example, the three identities $I=XX$, $I=YY$, and $I=ZZ$ are
replaced with the identity $I=AA$, and the three identities $Z_2 =
-X  Y$, $Y_2 = -ZX$, and $X_2 = -YZ$ are replaced with $A_2 =
-BC$.
\end{enumerate}

So roughly the filter applies known identities to try to shorten a
given one, and returns standardized, simplified (using certain
rules) identities.

\subsubsection{Filtering effects}

Table \ref{t:counts} shows the effects of filtering on the number
of identities returned. The first row lists the counts with no
filtering from section \ref{s:filter} applied. In this case, both
Mathematica 4.2 and C++ find 47 identities of length 1 (almost all
are trivial except $X_4=Y_4=Z_4=P_4=-I$). Going to length 2 added
625 identities, giving the total 672 in the table, and length 3
added 15068 identities, totaling 15740. The C++ program was fast
enough to find the length 4 identities, but the Mathematica
program is too slow for this length and longer. This case needed
to compute all $|\Lambda|^4=35^4=1,500,625$ combinations, and
compare them. Length 5 and 6 are probably computable, but will
take a lot of resources.

The second row of the table is the number of identities returned
when filtering is enabled, but not applying step \ref{i:rotation}
(rotation removal) and step \ref{i:grouping} (grouping similar
patterns) from section \ref{s:filter}. The third line contains the
counts when only step \ref{i:grouping} is skipped, and the last
line has counts obtained by applying all filtering steps.

\begin{table}[h]
\begin{center}
\begin{tabular}{|l||r|r|r|r|}\hline
Filter & Length 1 & Length $\leq2$ & Length $\leq3$ & Length $\leq4$ \\
\hline \hline

No filtering   & 47 & 672 & 15740 & 400089 \\ \hline 

Keep rots, No groups   & 12 &  66 &   293 &    1330 \\ \hline 

Drop rots, No groups    & 6  &  54 &   185 &   982 \\ \hline 

All filtering  & 2  &  36 &   155 &    931 \\ \hline 
\end{tabular}
\caption{Identity counts}\label{t:counts}
\end{center}
\end{table}

Looking at the table, and noting that even the 931 filtered length
4 identities are too tedious to put in a paper, I decided to put
the more manageable 155 length 3 and less identities. However the
unfiltered lists can be found online \cite{LomontCodeQC}.

\section{Identities}\label{s:identities}
The identities in table \ref{t:identities} are all the identities
resulting from 3 or fewer products of matrices from $\Lambda$ in
section \ref{s:definitions}, after filtering. They are sorted
alphabetically on the right hand side for quick reference. When
the symbols $A,B,$ or $C$ appear in the identities, that identity
stands for the three identities where $A,B,C$ are a cyclic
permutation of $X,Y,Z$, as explained in step \ref{i:grouping} in
section \ref{s:filter}.

\begin{table}[h]

{\footnotesize
\begin{align*}
I&=\;\;\;AA & Y&=\;\;\;HY_3Z_2 & Z_3&=-P_2Z_1Z & S&=\;\;\;X_2SY_2 & X&=\;\;\;Y_3HY_2\\
I&=-AC_2B & Y_1&=\;\;\;HZ & Z&=\;\;\;P_2Z_2 & Y&=\;\;\;X_2Y_3H & H&=\;\;\;Y_3HY_3\\
A_1&=-BA_1C & X_1&=\;\;\;HZ_1H & I&=\;\;\;P_2Z_2Z & H&=-X_2Y_3Y & S&=-Y_3SX_1\\
A_1&=-BA_3B & X_2&=\;\;\;HZ_2H & Z_1&=\;\;\;P_2Z_3Z & H&=-X_3HZ_1 & H&=-Y_3X\\
A_3&=\;\;\;BA_3C & Y_3&=\;\;\;HZ_2Y & Z_1&=-P_3S & S&=\;\;\;X_3SY_3 & X_2&=\;\;\;Y_3YH\\
A_2&=-BC & Y&=-HZ_2Y_1 & Z_2&=-P_3Z_1S & Y&=\;\;\;X_3YX_3 & H&=\;\;\;Y_3ZY_2\\
A&=\;\;\;BC_2 & X_3&=\;\;\;HZ_3H & Z_3&=-P_3Z_2S & Y&=-X_3ZX_1 & Y&=-Y_3Z_2H\\
A&=\;\;\;B_2C & Z_3&=-P_1ZS & I&=\;\;\;P_3Z_3S & Z&=\;\;\;X_3ZX_3 & H&=\;\;\;Y_3Z_2Y\\
A&=\;\;\;B_3AB_3 & S&=\;\;\;P_1Z_1 & S&=\;\;\;P_3Z_3Z & Y_1&=\;\;\;YHX_2 & Y_3&=-ZH\\
A&=\;\;\;B_3CB_1 & A_2&=-P_2A & A_4&=\;\;\;P_4 & Z_2&=-YHY_3 & X_1&=\;\;\;ZX_1Y\\
A_3&=-CA_3B & A_3&=-P_2A_1A & I&=\;\;\;P_4A_4 & Y_3&=\;\;\;YHZ_2 & X_3&=-ZX_1Z\\
A_1&=-CA_3C & A&=\;\;\;P_2A_2 & X_3&=-SHS & S&=\;\;\;YSX & Y_1&=-ZY_2H\\
A_2&=\;\;\;CB & I&=\;\;\;P_2A_2A & Z&=\;\;\;SS & X_3&=-YX_1Y & I&=-ZY_2X\\
A&=-CB_2 & A_1&=\;\;\;P_2A_3A & X_2&=-SXS & Y_3&=\;\;\;YX_2H & H&=-ZY_3\\
A&=\;\;\;C_1BC_3 & Y_3&=-P_2HX_2 & H&=\;\;\;SX_1S & H&=\;\;\;YX_2Y_3 & I&=-ZY_3H\\
A&=-C_2B & I&=\;\;\;P_2HX_2Y_1 & X&=\;\;\;SX_2S & I&=-YX_2Z & H&=-Z_1HX_3\\
A&=\;\;\;C_3AC_3 & X_2&=\;\;\;P_2HY_3 & Y_2&=-SYS & Z_1&=\;\;\;YZ_1X & X&=\;\;\;Z_1XZ_1\\
I&=\;\;\;HH & I&=-P_2HY_3X_2 & Y&=\;\;\;SY_2S & Z_3&=-YZ_1Y & Y&=-Z_1XZ_3\\
Y_3&=-HX & Y_1&=\;\;\;P_2HZ_2 & S&=\;\;\;TT & Z_1&=-YZ_3Y & Y&=\;\;\;Z_1YZ_1\\
Z_1&=\;\;\;HX_1H & I&=-P_2HZ_2Y_3 & Y_1&=\;\;\;XH & X&=\;\;\;Y_1H & H&=-Z_2HX_2\\
Z_2&=\;\;\;HX_2H & Y_2&=-P_2Y & H&=\;\;\;XY_1 & S&=-Y_1SX_3 & Y_1&=\;\;\;Z_2HY\\
Y&=-HX_2Y_3 & Y_3&=-P_2Y_1Y & Y_2&=-XY_3H & Y&=-Y_1X_2H & Y&=-Z_2HY_3\\
Z_3&=\;\;\;HX_3H & H&=\;\;\;P_2Y_1Z_2 & Z_3&=-XZ_1X & Z_2&=\;\;\;Y_1YH & Y_1&=-Z_2YH\\
Y_1&=-HYX_2 & I&=\;\;\;P_2Y_1Z_2H & Z_1&=-XZ_3X & H&=-Y_1YX_2 & H&=-Z_2Y_1Y\\
Z&=\;\;\;HY_1 & Y&=\;\;\;P_2Y_2 & Z_3&=\;\;\;XZ_3Y & H&=\;\;\;Y_1Z & H&=-Z_3HX_1\\
Y_2&=-HY_2H & I&=\;\;\;P_2Y_2Y & H&=-X_1HZ_3 & H&=\;\;\;Y_2HY_2 & X&=-Z_3YZ_1\\
Y_1&=-HY_2X & Z_2&=\;\;\;P_2Y_3H & S&=\;\;\;X_1SY_1 & Z&=\;\;\;Y_2HY_3 & Y&=\;\;\;Z_3YZ_3\\
X&=-HY_3 & I&=-P_2Y_3HZ_2 & Y&=\;\;\;X_1YX_1 & S&=-Y_2SX_2 & \\
Y_1&=-HY_3H & H&=-P_2Y_3X_2 & Z&=-X_1YX_3 & H&=\;\;\;Y_2XY_3 & \\
I&=-HY_3X & I&=-P_2Y_3X_2H & Z&=\;\;\;X_1ZX_1 & Z&=-Y_3H & \\
Z_2&=\;\;\;HY_3Y & Y_1&=\;\;\;P_2Y_3Y & Y_3&=\;\;\;X_2HY & Y&=-Y_3HX_2 & \\
Y_2&=-HY_3Z & Z_2&=-P_2Z & H&=-X_2HZ_2 & X_2&=-Y_3HY & \\
\end{align*}
}

\caption{The 155 filtered identities up to length
3}\label{t:identities}
\end{table}

\subsection{Hand simplification}
Simplifying diagrams by hand is quite tedious, but can be assisted
using these tables. Apply the procedure listed in section
\ref{s:procedure} to simplify the product until it simplifies no
further, then look for identities in the list that apply.

\subsubsection{Example 1} We verify the identity $Y=-XYX$. Starting
with $XYX$, we see no identities in the table starting with
$XY\dots$ or $YX\dots$, so we look for a grouping pattern, and
find $A_2=-BC$, which we apply as $Z_2=-XY$, giving
$XYX\rightarrow -Z_2X$. Expand using $A_2=CB$ and collapse $I=AA$
giving the complete transformation $XYX\rightarrow
-Z_2X\rightarrow -YXX\rightarrow -Y$, proving $Y=-XYX$.

\subsubsection{Example 2} Another way to approach this is to try
to get all expressions into $X,Y,Z,X_j,Y_j,Z_j$, and then commute
them to get $X$'s together, etc., until simplified. Thus to
simplify $HXH$, we have
$$
\labelrightarrow{HXH}{H=XY_1}{XY_1XXY_1}\labelrightarrow{}{I=AA}{XY_1Y_1}
\labelrightarrow{}{\text{step
\ref{i:collapse}}}{XY_2}\labelrightarrow{}{A=BC_2}{Z}
$$
where the rules applied are above each arrow, resulting in
$Z=HXH$.

\section{Conclusions and future work}\label{s:conclusion}
There are several ways to speed up the search process, like adding
the phase matrices only to the front, and not using the identity
except for comparisons. But the speed improvements are minimal.

Other directions are to extend this to understanding the 2 and 3
qubit operations. For example, the Toffoli gate can be implemented
with 5 basic 2 qubit gates \cite{BBCD95}. I believe it is unknown
if this can be done with 4 gates, although it seems unlikely. A
computer search should be able to shed light on this, and perhaps
open up new understanding about minimal number of quantum gates
needed for some other constructions.

Finally, length 2, 3, and 4 filtered and unfiltered identities are
online \cite{LomontCodeQC}, as well as the C++ code and the
Mathematica 4.2 code.

\bibliographystyle{amsplain}
\bibliography{qbib}
\end{document}